\documentstyle[prl,aps,epsf,multicol,tabularx]{revtex} 
\tighten 
\begin{document} 
\draft 
\title{CHARGE FLUCTUATIONS IN A QUANTUM POINT CONTACT ATTACHED TO A 
SUPERCONDUCTING LEAD} 
\author{Andrew M. Martin, Thomas Gramespacher and Markus B\"uttiker} 
\address{D\'epartement de Physique Th\'eorique, Universit\'e de Gen\`eve, 
CH-1211 Gen\`eve 4, Switzerland.} 
\date{\today} 
\maketitle 
\begin{abstract} 
We show how to calculate the charge noise spectrum in a normal 
mesoscopic conductor, which is capacitively coupled to a 
macroscopic gate, when this conductor is attached to $L$ normal 
leads and $M$ superconducting leads, the only restriction being 
that the superconducting leads must be at the same chemical 
potential. We then proceed to examine results for a quantum 
point contact (QPC) in a normal lead connecting to a 
superconductor. Of interest is the 
fluctuating current in a gate capacitively coupled to a QPC. 
The results are compared with the case when all leads are normal. 
We find a doubling of the equilibrium charge fluctuations and a 
large enhancement ($>2$) in the current noise spectrum to first order in 
$|eV|$, when a channel in the QPC is 
opening. 
\end{abstract} 
\begin{multicols}{2} 
\narrowtext 
 
\begin{figure} 
\narrowtext 
\vspace*{-0.0cm} 
\epsfxsize=7cm \centerline{\epsffile{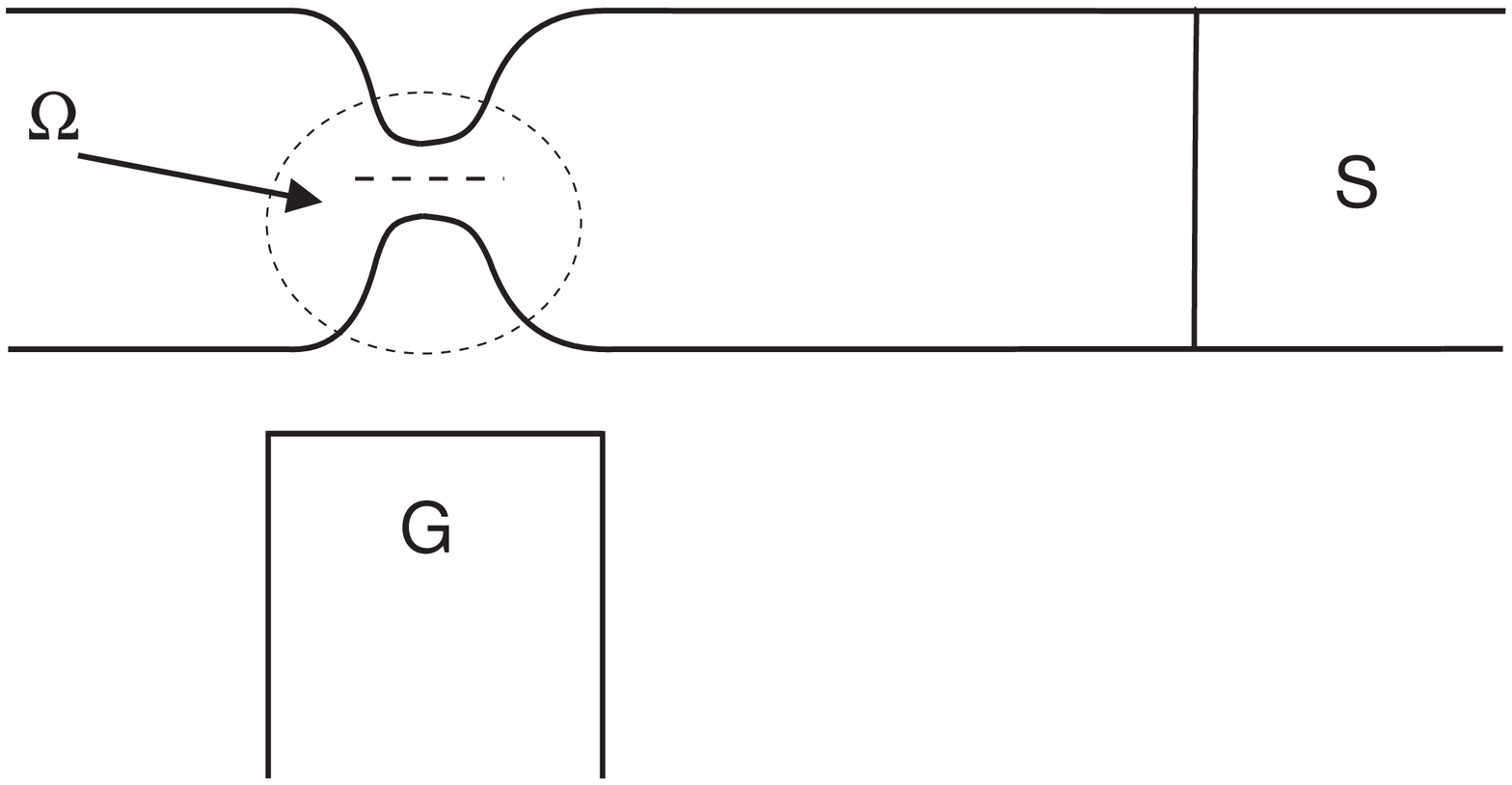}} \vspace*{0.5cm} 
\caption{Quantum point contact, attached to one normal and one 
superconducting lead, capacitively coupled to a gate.} 
\label{Fig1} 
\end{figure}

Theory and experimental measurements of the electrical transport 
properties of mesoscopic conductors, which either contain 
superconducting regions or are attached to superconducting leads 
have generated great interest. Of particular interest are the 
noise properties of normal superconducting interfaces. While many 
aspects of the low frequency-current noise spectra \cite{Khlus} 
have been understood by generalizing the scattering approach 
\cite{deJong,Muzy,Datta,Beenakker,super} of normal conductors, the 
fluctuations of the charge have remained unexplored. We generalize 
the work done for normal mesoscopic 
conductors\cite{Buttiker93,Buttiker96,Pedersen} to systems which 
contain superconducting leads. The only restrictions are that 
there must be one or more normal leads, the superconducting leads 
must all be at the same chemical potential and any gates in the 
structure only {\it see} normal regions of the conductor. Having 
developed this technique we shall consider one particular example 
shown in Fig. \ref{Fig1}, a Quantum Point Contact (QPC) attached 
to one normal lead, one superconducting lead and capacitively 
coupled, via the Coulomb interaction, to a macroscopic gate. We 
are interested in the charge fluctuations in the hybrid structure 
which can be measured by observing the current fluctuations at the 
gate. It turns out that at equilibrium the current fluctuations 
are determined by the $RC$ -time constant. Thus our primary aim is 
to find the charge relaxation resistance $R$ and the capacitance 
$C$ for hybrid 
\noindent 
structures. In the presence of transport, 
a novel 
resistance $R_V$, which reflects 
the shot noise \cite{Khlus,deJong,Muzy} of the hybrid structure, appears.

Consider a mesoscopic conductor with 
$L$ normal ideal leads and $M$ superconducting leads. 
The conductor can be described by a scattering matrix 
with elements $s_{\gamma_\alpha 
\delta_\beta}$ which relate particle and hole amplitudes 
($\alpha, \beta = p,h$) 
incoming at contact $\delta$ to the outgoing amplitudes at 
contact $\gamma$. Over a 
region $\Omega$ the 
mesoscopic conductor is capacitively 
coupled to a gate. The current fluctuations induced into 
the gate can be found from the charge fluctuations of the mesoscopic 
conductor\cite{Buttiker96,Pedersen}. The investigation of the charge 
fluctuations starts from the analysis of the bare charge fluctuations 
which are then screened to find the 
true charge fluctuations \cite{Buttiker96,Pedersen}. 
For a normal conductor the bare charge fluctuations 
are determined by a density of states matrix which has dimensions $L*L$. 
For hybrid structure the matrix has dimensions $8*L*L$: 
Any scattering channel, independently on whether the incoming 
or outgoing channels are electron or hole like, can contribute 
to the electron density at a point ${\bf r}$ inside the conductor. 
Thus to express the density in terms of the scattering matrix 
requires a conceptual trick: It is useful to imagine that there 
are two electrostatic potentials $U^{p}$ and $U^{h}$ which act each 
separately on the electrons and holes \cite{Thomas99}. 
The matrix which governs the charge fluctuations can then be found 
by testing the scattering matrix elements with regard 
to small variations of the electron and hole 
potentials. This procedure gives the following density 
of states elements \cite{Thomas99} 
\begin{eqnarray} 
{\cal N}_{\gamma_\alpha \delta_\beta}^{\eta}(\nu_\lambda, {\bf r}) 
&=& \frac{-1} {4\pi i}  \left[ 
s_{\nu_\lambda\gamma_\alpha}^{\dagger}(E,\underline{U}({\bf 
r})) \frac{ 
\partial s_{\nu_\lambda \delta_\beta}(E,\underline{U}({\bf r}))} 
{q^{\eta} \partial U^{\eta}({\bf r})} \right. \nonumber \\ & - & 
\left. s_{\nu_\lambda\gamma_\alpha}(E,\underline{U}({\bf 
r})) \frac{ 
\partial s_{\nu_\lambda \delta_\beta}^{\dagger}(E,\underline{U}({\bf r}))} 
{q^{\eta} \partial U^{\eta}({\bf r})} \right] \label{pdos} 
\end{eqnarray} 
where the labels $\nu$, $\gamma$ and $\delta$ denote contacts 
($1....L$), $\alpha$, $\beta$, $\lambda$ and $\eta$ denote the 
electron/hole degrees of freedom ($p/h$) and $q^{p}=e=-q^{h}$. The 
functional derivatives are taken at the equilibrium electrostatic 
potential $U^{p} = U^{h} = U_{eq}$. To give an example, ${\cal 
N}_{1p 2p}^{h}(1_h, {\bf r})$ is the hole density associated with 
two electron current amplitudes incident from contacts $1$ and $2$ 
and a reflected outgoing hole amplitude in contact $1$. With the 
help of these basic expressions we can now find both the average 
density of states as well as the fluctuations. The hole density of 
states of a region $\Omega$ of the conductor is 
$N^h=N^h(p)+N^h(h)$ where 
\begin{equation} 
N^h(\alpha)= \sum_{\nu \gamma \lambda}\int_{\Omega}d^3{\bf r} 
{\rm Tr} [{\cal N}_{\gamma_\alpha \gamma_\alpha}^{h}(\nu_{\lambda},{\bf 
r})] \label{hdos} 
\end{equation} 
and the particle density of states is $N^p=N^p(p)+N^p(h)$ with 
\begin{equation} 
N^p(\alpha)= \sum_{\nu \gamma \lambda} \int_{\Omega}d^3 {\bf r} 
{\rm Tr} [{\cal N}_{\gamma_\alpha \gamma_\alpha}^p(\nu_{\lambda}{\bf r})]. 
\label{ppdos} 
\end{equation} 
The trace is over open quantum channels. 
$N^{\alpha}(\beta)$ is the injectivity of particles (holes) ($\beta=p (h)$), 
from all contacts into the conductor, given a change in the particle (hole) 
potential ($\alpha=p(h)$). 
 
The fluctuations of the bare charge 
in a region $\Omega$ of interest 
can be found from the charge operator 
$e\hat{{\cal N}}$ given by 
\begin{eqnarray} 
& & e \hat{{\cal N}} (\omega) = 
\sum_{\gamma \alpha \atop \delta \beta} \sum_{\eta \nu \lambda}
\int_{\Omega} d^3 {\bf r} 
\int dE \hat{a}^{\dagger}_{\gamma_\alpha} (E)  \nonumber \\ 
& & \times q^{\eta}
{\cal N}^{\eta}_{\gamma_\alpha \delta_\beta} 
(\nu_{\lambda}, {\bf r}; E, E+\hbar \omega) 
\hat{a}_{\delta_\beta} (E+\hbar \omega) 
\label{nfluct} 
\end{eqnarray} 
where the zero-frequency 
limit of 
${\cal N}^{\eta}_{\gamma_\alpha \delta_\beta} 
(\nu_{\lambda}, {\bf r}; E, E+\hbar \omega)$ 
is given by Eq. (\ref{pdos}). 
In Eq. (\ref{nfluct}), 
$\hat{a}^{\dagger}_{\gamma_\alpha} (E)$ creates an incoming 
electron/hole ($\alpha=p/h$) in lead $\gamma$. 
The true charge fluctuations must be obtained by taking into 
account the Coulomb interaction and below we show how to obtain 
the true charge fluctuations from the fluctuations of the bare 
charges. 
 
Given equations (\ref{pdos}-\ref{nfluct}) our next step is to 
consider the system shown in Fig. 1 and to calculate the charge 
fluctuations in the mesoscopic region and hence deduce the current 
fluctuations in the gate. To do this it helps to first consider an 
approach which allows us to correctly calculate the charge $\delta 
Q$ in the mesoscopic region when a voltage is applied across the 
system. The charge accumulated underneath the gate in the region 
$\Omega$ can be expressed in two ways 
\cite{Buttiker93,Buttiker96,Pedersen}: If we describe the Coulomb 
interaction with the help of a geometrical capacitance $C$ the 
incremental charge is simply related to the potential variations 
in the conductor $\delta U$ and the gate $\delta V_g$ via $\delta 
Q  =  C(\delta U-\delta V_g)$. On the other hand, the charge 
$\delta Q$ is also the sum of the injected charges due to the 
variation of the contact potentials (keeping the internal 
electrostatic potential fixed) and the induced charges which are 
generated by the Coulomb interaction. Thus 
\begin{eqnarray} 
\delta Q  &=&  C(\delta U-\delta V_g) = 
 \frac{1}{2} \left[q^p N^p(p)\delta \mu_p + q^h N^h(p)\delta \mu_p \right. 
 \nonumber \\ 
  &+& q^h N^h(h)\delta \mu_h + q^p N^p(h)\delta \mu_h \nonumber \\ 
   &-&q^p N^p(p) q^p \delta U_p- q^p N^p(h) q^h \delta U_h \nonumber \\ 
  &-& q^h N^h(p) q^p \delta U_p - \left. q^h N^h(h) q^h \delta U_h \right] 
\label{charge} 
\end{eqnarray} 
The first four terms in Eq. (\ref{charge}) determine the charge 
{\it injected} into the conductor as a consequence of the 
variation of the contact potentials. Similarly to the conceptual 
separation of the electrostatic potential, we have here assumed 
that each contact has separate electrochemical potentials $\mu_p$ 
and $\mu_h$ for electrons and holes \cite{Datta}. In reality 
$\mu_p = - \mu_h = \mu= eV$. The second four terms in Eq. 
(\ref{charge}) determine the {\it induced} charges. In the absence 
of superconductivity Eq. (\ref{charge}) reduces to the "Poisson 
equation" for charge variations in a normal conductor 
\cite{Buttiker93,Pedersen}: We have $N^p(h) = N^h(p) = 0$ and $N 
\equiv N^p(p) = N^h(h)$. Going back to Eq. (\ref{charge}) we can 
solve it for $\delta U$ and can use this solution to determine the 
electrochemical capacitance of the hybrid structure vis-a-vis the 
gate. We find 
\begin{equation} 
C^{NS}_{\mu} \equiv \frac{\delta Q}{\delta V } \equiv - 
\frac{\delta Q}{\delta V_{g} } 
= \frac{C e^2N_{\Sigma}}{C+e^2N_{\Sigma}} 
\end{equation} 
with a total density of states 
\begin{equation} 
N_{\Sigma}=\frac{1}{2}[N^p(p)-N^p(h)+N^h(h)-N^h(p)]. 
\end{equation} 
It is instructive to examine 
$C^{NS}_{\mu}$ for an ideal ballistic 
wire. For a perfect $N-S$ structure, $C_{\mu}$ is zero, 
since every charge incident upon the conductor is perfectly Andreev 
reflected and hence the net accumulated charge is zero. If the superconductor 
is driven into the normal state, the capacitance \cite{Buttiker93,Guo} 
is $C^{N}_{\mu} 
= {C e^2N}/({C+e^{2} N})$ which for a ballistic wire reduces to 
$C^{N}_{\mu} = C$ since typically $e^{2}/C >> 1/N$. Thus $C^{NS}_{\mu}$ 
for the hybrid structure can differ dramatically 
from $C^{N}_{\mu}$. 
 
If we now wish to consider fluctuations in the charge 
then we have to consider the Poisson equation 
for the fluctuating charges \cite{Pedersen,Buttiker96}. 
For the case that the voltages are held fixed 
(zero-impedance external circuit) 
this leads to an operator equation, 
\begin{equation} 
\hat{Q} = C \hat{U} = e \hat{{\cal N}} - e^{2} N_{\Sigma} \hat U 
\label{pfluct} 
\end{equation} 
where $\hat{{\cal N}}$ is the operator of 
the bare charge fluctuations given by Eq. (\ref{nfluct}) 
and the last term in Eq. (\ref{pfluct}) describes the 
screening of the bare charge fluctuations. 
Solving Eq. (\ref{pfluct}) for ${\hat U}$ we can express 
the charge fluctuations ${\hat Q}$ in terms of the bare 
charge fluctuations \cite{Pedersen}. 
For the fluctuation spectrum of the charge this gives 
\begin{eqnarray} 
& &S_{QQ}(\omega)  = (1/2) 
C_{\mu}^2 N_{\Sigma}^{-2} 
\sum_{\gamma \delta} 
\sum_{\alpha \beta} 
\int 
dE F_{\gamma_{\alpha} \delta_{\beta}}(E,\hbar \omega)  \nonumber \\ 
& &\times {\rm Tr} 
[{\cal N}_{\gamma_{\alpha} \delta_{\beta}}(E, E+\hbar \omega) 
{\cal N}_{\gamma_{\alpha} \delta_{\beta}}^{\dagger}(E, E+\hbar 
\omega)] 
\label{eq:fluct} 
\end{eqnarray} 
where 
\begin{eqnarray}
& &{\cal N}_{\gamma_\alpha \delta_\beta} (E, E^{\prime})=
\sum_{\eta\nu\lambda} \! {\rm sgn} (q^{\eta})\!\!\! \int_{\Omega}\!\!\! d^3 {\bf r} 
{\cal N}^{\eta}_{\gamma_\alpha \delta_\beta} 
(\nu_{\lambda}, {\bf r}; E, E^{\prime}),\\
& & F_{\gamma_{\alpha} \delta_{\beta}}(E, E^\prime) =  
f_{\gamma_{\alpha}}(E)[1-f_{\delta_{\beta}}(E^\prime)] \nonumber \\ 
& & \qquad + f_{\delta_{\beta}}(E^\prime)[1-f_{\gamma_{\alpha}}(E)]. 
\end{eqnarray} 
In the above the sum is over all normal contacts $\gamma \delta$ 
and degrees of freedom $\alpha \beta$. The Fermi functions 
$f_{\gamma_{\alpha}}(E)$ are defined such that 
$f_{\gamma_{p}}=f_0(E_p-\mu_{\gamma})$ and 
$f_{\gamma_{h}}=f_0(E_h+\mu_{\gamma})$ where $E_{\alpha}$ is the 
energy of a particle (hole) ($\alpha= p(h)$) in reservoir 
$\gamma$, which is at a chemical potential $\mu_{\gamma}$, 
$f_0(E)$ is the Fermi function at the condensate chemical 
potential of the superconducting leads ($\mu_0$). 
 
Now we evaluate Eq. (\ref{eq:fluct}) at equilibrium and zero 
temperature, to leading order in $\hbar \omega \ll |\Delta|$. We find 
\begin{equation} 
S_{QQ}(\omega)=2 C_{\mu}^2 R_q \hbar |\omega| 
\label{sqqo} 
\end{equation} 
with a charge relaxation resistance 
\begin{equation} 
R_q=\frac{h}{4e^2}\frac{\sum_{\gamma \delta} \sum_{\alpha \beta} 
{\rm Tr} ({\cal N}_{\gamma_{\alpha} \delta_{\beta}} {\cal N}_{\gamma_{\alpha} 
\delta_{\beta}}^{\dagger})}{[N_{\Sigma}]^2}. 
\label{rq} 
\end{equation} 
In the zero frequency limit, and at a temperature $kT \ll |\Delta|$ 
the charge fluctuations are given by $S_{QQ}(\omega)=2 
C_{\mu}^2 R_q  kT $ with $R_q$ as given by Eq. (\ref{rq}). If 
we bring the hybrid structure into a non-equilibrium state by 
applying a bias $e|V| \gg kT$ between the normal reservoir and the 
superconductor, we find at zero temperature in the low frequency 
limit to leading order in $e|V|$ for the system shown in Fig. 
\ref{Fig1}, 
\begin{equation} 
S_{QQ}(\omega)=2C_{\mu}^2 R_V  e |V| 
\label{sqqv} 
\end{equation} 
with a non-equilibrium resistance 
\begin{equation} 
R_V=\frac{h}{4e^2}\frac{ {\rm Tr }(  
{\cal N}_{1_{p}1_{h}} {\cal N}_{1_{p}1_{h}}^{\dagger}+ 
{\cal N}_{1_{h}1_{p}} {\cal N}_{1_{h}1_{p}}^{\dagger})}{[N_{\Sigma}]^2}. 
\label{nrq} 
\end{equation} 
The resistance $R_V$ is a consequence of the charge fluctuations 
which arise due to the {\it shot} noise of the transport state. 
The fluctuation spectra of the current at the gate $G$ in Fig. 
(\ref{Fig1}) are related to the charge fluctuation spectra Eqs. 
(\ref{sqqo}) and (\ref{sqqv}) via $S_{II}(\omega) = 
\omega^{2} S_{QQ}(\omega)$.

To proceed further we follow Beenakker \cite{Beenakker} 
and express the electron-hole scattering matrix of the NS-structure 
in terms of the scattering matrix elements of the normal structure. 
For simplicity we restrict our considerations here to normal 
structures which are symmetric with respect 
to left and right going carriers. The scattering 
matrix of the normal conductor is then given 
by a reflection matrix $r$ and a transmission matrix $t$ only. 
In terms of $r$ and $t$ the electron-hole scattering 
matrix elements are 
\begin{eqnarray} 
s_{1_p1_p}(E)&=&r_p(E) + \alpha^2 t_p(E) r_h(-E) M_p(E) t_p(E) 
\label{beenak1} \\ 
s_{1_h1_p}(E)&=&\alpha e^{-i \phi} t_h(-E) M_p(E) t_p(E) 
\label{beenak2}\\ 
s_{1_h1_h}(E)&=&r_h(E)+ \alpha^2 t_h(E)r_p(-E)M_h(E) t_h(-E) 
\label{beenak3}\\ 
s_{1_p1_h}(E)&=&\alpha e^{i \phi} t_h(E) M_h(E) t_p(-E) 
\label{beenak4} 
\end{eqnarray} 
where $\alpha=\exp[-i\rm{arccos}(E/\Delta)]$ and 
\begin{eqnarray} 
M_p(E) &=&  \left[1-\alpha^2 r_p(E) r_h(-E) \right]^{-1} \\
M_h(E) &=&  \left[1-\alpha^2 r_h(E) r_p(-E) \right]^{-1}. 
\label{beenaklast}
\end{eqnarray} 
In Eqs. (\ref{beenak1}-\ref{beenaklast}) 
we have given an index $p,h$ to $r$ and $t$ to indicate 
whether a carrier or a hole is scattered since we have 
to distinguish derivatives 
with respect to  the electron and hole potentials. 
Throughout the rest of the work presented here we will be making 
use of particle-hole symmetry (ie. $E \rightarrow 0$ $s_p = 
s_h^{\star}$). 
 
Now we wish to calculate the properties of the system shown in Fig. 1. We 
choose to model a quantum point contact (QPC), using a saddle point 
potential \cite{Buttiker90} 
$V(x,y)=V_0+\frac{1}{2}m\omega_y^2y^2-\frac{1}{2}m\omega_x^2x^2 $ 
and as in Ref.\ \cite{Pedersen} we have evaluated the density of 
states semi-classically. We also take the WKB limit allowing us 
to make the following transformation \cite{Buttiker96} 
$- \int_{\Omega} d^{3}{\bf r} 
\frac{\partial}{\partial  q^{\lambda} U^{\lambda}({\bf r})} 
\rightarrow d/dE^{\lambda} $. 
In the above transformation we have been careful to keep the index 
$\lambda$ included, such that we can distinguish between particles 
and holes. We also note that we can write 
$ds/dE^{\lambda}= 
(ds/d\theta^{\lambda}) (d\theta^{\lambda}/dE^{\lambda})$ 
where $d\theta^{\lambda}/dE^{\lambda}$ is the particle (hole) 
($\lambda=p(h)$) density of states divided by $\pi$ in the conductor 
in region $\Omega$. Making use of 
particle-hole symmetry we have 
$d\theta/dE=d\theta^p/dE^p=d\theta^h/dE^h $.

\begin{figure}
\narrowtext 
\vspace*{0.0cm} 
\epsfxsize=7cm \centerline{\epsffile{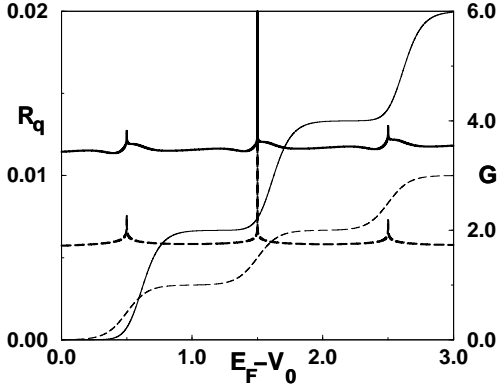}} 
%TOM\vspace*{0.5cm} 
\caption{ \label{Fig3} $R_q^{NS}$ (thick solid line) and $R_q^{N}$ 
(thick dashed line), in units of $h/e^2$. $G^{NS}$ 
(thin solid line) and $G^N$ (thin dashed line) is the conductance, in units 
of $e^2/h$, as a function of $E_F-V_0$ in units of $\hbar \omega_y$, for a 
saddle QPC with 
$\omega_y/\omega_x=2$. } 
\end{figure}

Using Eqs. (\ref{rq},\ref{beenak1}-\ref{beenak4}) we now calculate $R_q$ 
and find 
\begin{equation} 
R_q^{NS}=\frac{h}{2e^2} 
\frac{\sum_n \frac{16R^6_n+R_n(4R_n+1)^2T_n^2}{(1+R_n)^4} 
\left(\frac{d \theta_n}{dE}\right)^2} 
{\left[\sum_n \frac{4R_n^3}{(1+R_n)^2} 
\left(\frac{d \theta_n}{dE}\right)\right]^2} 
\end{equation} 
as compared \cite{Pedersen,note} to 
\begin{equation} 
R_q^{N}=\frac{h}{4e^2} 
\frac{\sum_n \left(\frac{d \theta_n}{dE}\right)^2} 
{\left[\sum_n \left(\frac{d \theta_n}{dE}\right)\right]^2} 
\end{equation} 
for a QPC attached to two normal leads, where $T_n$ and $R_n=1-T_n$ are the 
transmission and reflection probabilities of the $n$th quantum channel and 
$\theta_n$ is the phase accumalated by a carrier in the $n$th channel during 
transmission through the QPC. In Fig. \ref{Fig3} we plot $R_q^{NS}$ and 
$R_q^{N}$ (the charge 
relaxation resistance 
for a QPC attached to two normal leads). For 
the parameters we have chosen, we see that the 
difference between $R_q^N$ and $R_q^{NS}$ is roughly 
a factor $2$. This arises from the difference between the contact 
resistance of a mesoscopic conductor in an N-S system compared to an 
N-N system \cite{sols}. 

\begin{figure}
\narrowtext \vspace*{0.0cm} 
\epsfxsize=7cm \centerline{\epsffile{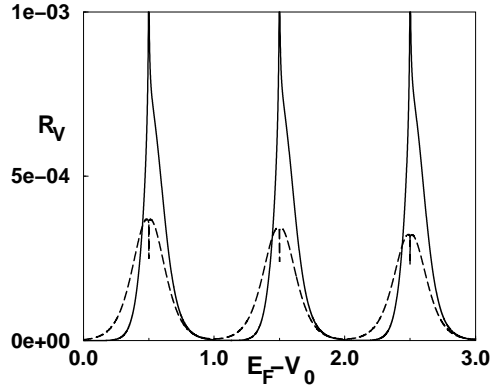}} 
%TOM\vspace*{0.5cm} 
\caption{ \label{Fig4} $R_V^{NS}$ (solid line) and $R_V^{N}$ 
(dashed line), in units of $h/e^2$,  as a function of $E_F-V_0$ 
in units of $\hbar \omega_y$, 
for a saddle QPC with $\omega_y/\omega_x=2$. } 
\end{figure}

Having examined $R_q^{NS}$ and $R_q^{N}$ we now proceed to examine 
$R_V^{NS}$, finding
\begin{equation} 
R_V^{NS}= 
 \frac{h}{2e^2} 
\frac{\sum_n \frac{R_n(4R_n+1)^2T_n^2}{(1+R_n)^4} 
\left(\frac{d \theta_n}{dE}\right)^2} 
{\left[ \sum_n \frac{4R_n^3}{(1+R_n)^2} 
\left(\frac{d \theta_n}{dE}\right)\right]^2} 
\end{equation} 
%TOMas compared to
% 
%\noindent plot $R_q^{NS}$ and $R_q^{N}$ (the charge 
%relaxation resistance 
%for a QPC attached to two normal leads). For 
%the parameters we have chosen, we see that the 
%difference between $R_q^N$ and $R_q^{NS}$ is roughly 
%a factor $2$. 
% 
%Having examined $R_q^{NS}$ and $R_q^{N}$ we now proceed to examine 
%$R_V^{NS}$, finding 
%\begin{equation} 
%R_V^{NS}= 
% \frac{h}{2e^2} 
%\frac{\sum_n \frac{R_n(4R_n+1)^2T_n^2}{(1+R_n)^4} 
%\left(\frac{d \theta_n}{dE}\right)^2} 
%{\left[ \sum_n \frac{4R_n^3}{(1+R_n)^2} 
%\left(\frac{d \theta_n}{dE}\right)\right]^2}. 
%\end{equation} 
%TOM
\noindent
as compared \cite{Pedersen} to 
\begin{equation} 
R_V^{N}= 
 \frac{h}{e^2} 
\frac{\sum_n  \frac{1}{4R_nT_n}\left(\frac{d T_n}{dE}\right)^2} 
{\left[ \sum_n \left(\frac{d \theta_n}{dE}\right)\right]^2} 
\end{equation} 
for a QPC attached to two normal leads. 
In Fig. \ref{Fig4} we see both $R_V^{NS}$ and $R_V^{N}$ 
as a function of $E_F-V_0$. At 
the channel threshold $R_n = T_n = 1/2$ the non-equilibrium 
resistance $R_V^{NS}$ of the hybrid structure tends to $(9/4) 
(h/e^2)$ whereas the normal state $R^{NS}_V$ is completely 
suppressed at the channel threshold due to the divergence of the 
semiclassical density of states.

In this work we have taken the first steps in considering 
frequency-dependent charge fluctuations and noise 
calculations in mesoscopic normal superconducting hybrid systems. 
To do this we have considered structures where the condensate 
chemical potentials of any superconducting lead remains the same 
and gates only {\it see} the normal regions of the conductor. 
This allows us to consider oscillating voltages in the normal 
reservoirs and calculate the systems response. We have focused on the 
$RC$-time of such structures. The $RC$-time is a fundamental 
dynamical quantity of electrical conductors. Elsewhere we have 
shown that for normal mesoscopic conductors the $RC$-time also 
determines the dephasing rate of Coulomb coupled structures\cite{mbam}. 
We expect that such a relation also holds 
for hybrid systems. As an example we considered the most 
simple non-trivial example, a QPC attached to one normal lead and 
one superconducting lead. We have given a firm prediction that the 
$RC$-time and the fluctuations induced into a gate differ markedly 
for an NS structure and a normal conductor. As usual, we assume 
perfect Andreev reflection at the superconducting interface, 
whereas at real interfaces only partial Andreev reflection might 
occur. Currents induced into gates can be measured \cite{Chen}. 
We believe that it is similarly possible to measure the current 
fluctuations at a gate.
 
This work was supported by the Swiss National Science Foundation 
and by the TMR network Dynamics of Nanostructures.

\end{multicols} 

\begin{references}
\bibitem{note} The factor $1/4$ was mistakenly omitted in $R_{q}$ 
as given in Ref. \cite{Pedersen} and is herewith corrected.
\end{references}
\end{document}